\documentclass[a4paper,11pt]{article}
\usepackage{jheppub}
\usepackage[T1]{fontenc}
\usepackage{amsfonts}
\usepackage{amsmath}
\usepackage{amssymb,epsf}
\usepackage{latexsym}
\usepackage{graphicx,epsfig}
\usepackage{amssymb}
\usepackage{subfigure}

\title{Analytical study of properties of holographic superconductors with exponential nonlinear electrodynamics}
\author[a,b]{A. Sheykhi}
\author[a]{F. Shaker}
\affiliation[a]{ Physics Department and Biruni Observatory,
College of Sciences, Shiraz University, Shiraz 71454, Iran}
\affiliation[b]{Research Institute for Astronomy and Astrophysics
of Maragha (RIAAM), P.O. Box 55134-441, Maragha, Iran}
\emailAdd{asheykhi@shirazu.ac.ir}

\abstract{Based on the Sturm-Liouville (SL) eigenvalue problem, we
analytically study several properties of holographic $s$-wave
superconductors with exponential nonlinear electrodynamics in the
background of Schwarzschild anti-de Sitter (AdS) black holes. We
assume the probe limit in which the scalar and gauge fields do not
back react on the background metric. We show that for this system,
one can still obtain an analytical relation between the critical
temperature and the charge density. Interestingly enough, we find
that logarithmic nonlinear electrodynamics decreases the critical
temperature, $T_c$, of the holographic superconductors compared to
the linear Maxwell field. This implies that the nonlinear
electrodynamics make the condensation harder. The analytical
results obtained in this paper are in good agreement with the
existing numerical results. We also compute the critical exponent
near the critical temperature and find out that it is still $1/2$
which seems to be an universal value in mean field theory.}

\begin{document}
\maketitle \flushbottom

\section{Introduction}
Nowadays, the correspondence between an anti-de Sitter spacetime
and a conformal field theory living on its boundary (AdS/CFT) is
the well-known holographic picture of the world. This
correspondence provides a method for relating a strongly
interacting field theory to a dual classical gravity
\cite{cor1,cor2,cor3}. It was shown that by employing the
dictionary of  AdS/CFT duality, one can relate the observed
properties in the AdS bulk to their corresponding quantities in
CFT on the boundary. An interesting application of this duality
was recently proposed by Hartnoll, et. al., \cite{Har1,Har2} who
showed that some properties of strongly coupled superconductors
can be potentially described by a classical general relativity
living in one higher dimension. The motivation for this study is
to shed some light on the problem of understanding the mechanism
of the high temperature superconductors in condensed matter
physics. Following \cite{Har1,Har2}, a lot of works have been
appeared which attempt to investigate the holographic
superconductors from different perspective \cite{sup1, sup2, sup3,
sup4, sup5, sup6, sup7, sup8, sup9,XH, RGC2,Wang2,Pin, Zeng}.
Focusing on the simplest form by working in the probe
approximation, in which the scalar and gauge fields backreaction
effects on the gravity sector can be ignored and fixing the metric
as static, the authors of \cite{P.ST2010} employed the SL
variational method as an analytic approach for calculating the
properties of the holographic superconductors. They discussed
analytic approximations to the solutions of the nonlinear field
equations both near the critical temperature $T_{c}$ and in the
low temperature limit $T \rightarrow 0$. The critical temperature
and critical exponent and other quantities for different modes of
superconductors including $s$-wave, $p$-wave and $d$-wave, were
also investigated in the probe limit \cite{P.GR,P.BGRL,P.CW,
P.JPC, P.GTW}.

In description of the holographic dual models, some attempts have
been done to generalize the linear Maxwell electrodynamics to the
nonlinear electrodynamics. The motivation is to examine the
effects of the higher order corrections to the Maxwell field on
the properties and critical temperature of the holographic
superconductors. In this regards, the authors of
\cite{BIGB,LPJW2015} analytically studied the holographic
superconductors in the presence of the Born-Infeld (BI) nonlinear
electrodynamics by using the SL eigenvalue problem. They observed
that the higher BI corrections make it harder for the condensation
to form but do not affect the critical phenomena of the system.
Also, the authors of \cite{LGW.NN} numerically studied the
non-equilibrium condensation process in a holographic
superconductor when the gauge field is in form of the BI and
Logarithmic one.

Among the nonlinear corrections to the linear Maxwell field, it is
of interest to consider the exponential form of the nonlinear
electrodynamics \cite{Hendi}. Numerical studies on the properties
of the holographic superconductors in the presence of BI nonlinear
electrodynamics \cite{JC} as well as logarithmic and exponential
nonlinear (EN) electrodynamics have been carried out
\cite{ZPCJ.NN,JC}. In particular, it was shown that EN
electrodynamics correction makes the condensation harder
\cite{ZPCJ.NN}. When the gauge field is the BI one, based on the
SL eigenvalue problem, several properties of holographic $s$-wave
superconductors in the background of a Schwarzschild-AdS spacetime
have been investigated in the prob limit \cite{Gan,GanGB} and away
from the probe limit \cite{BIBR,BI-GB-BR}. As far as we know the
analytical study on the properties of holographic superconductors
with EN electrodynamics remains to be done. In this paper, we
would like to extend the analytical study on the holographic
$s$-wave superconductors by taking the gauge field in the form of
EN electrodynamics and using SL analytical techniques. For
economic reasons, we shall concentrate on the probe limit by
neglecting  backreaction of the matter fields on the spacetime
metric. We shall compute the critical temperature and critical
exponent for this kind of holographic superconductor by using the
perturbation method. Then, we will compare our results with the
existing numerical results for holographic superconductors with EN
electrodynamics \cite{ZPCJ.NN} as well as the analytical results
derived in \cite{Gan} for the holographic superconductors with BI
nonlinear electrodynamics.

The plan of this paper is as follows. In section \ref{Basic}, we
review the field equations and study the $(2+1)$-holographic
superconductor in the framework of planar AdS black holes when the
gauge field is in the form of EN electrodynamics. In section
\ref{Field}, we calculate the critical temperature in terms of the
charge density. In section \ref{Cri}, we continue the calculation
for the condensations of the scalar operators and find an
analytical expression for it. We show that the analytic study is
comparable with the numerical computation. Finally, section
\ref{Con} contains our concluding remarks.
\section{Basic Equations \label{Basic}}
We study the formation of scalar hair on the background of AdS
black hole in $4$-dimensions. In order to construct a holographic
$s$-wave superconductor in the probe limit, we consider the
background of the $4$-dimensional planar Schwarzschild-AdS black
hole
\begin{eqnarray}
ds^2=-f(r) dt^2+\frac{1}{f(r)}dr^2+r^2(dx^2+dy^2),
\end{eqnarray}
where the metric function is
\begin{eqnarray}
f(r)=r^2-\frac{r_{+}^3}{r}.
\end{eqnarray}
Here $r_{+}$ is the horizon radius of the black hole and we have
taken the AdS radius equal unity, i.e. $l=1$. The Hawking
temperature of the Schwarzschild-AdS black hole can be expressed
as
\begin{eqnarray}\label{T}
T=\frac{3r_{+}}{4\pi}.
\end{eqnarray}
The Lagrangian density containing a nonlinear gauge field and  a
charged complex scalar field reads
\begin{equation}\label{L}
\mathcal{L}=\mathcal{L}(F)-| \nabla\psi - i q A \psi|^{2} - m^{2}
|\psi|^{2},
\end{equation}
where  $A$ and $\psi$ are, respectively, the gauge and scalar
field with charge $q$, $\mathcal{L}(F)$ denotes the Lagrangian
density of the EN electrodynamics and we take it as \cite{Hendi}
 \begin{equation}
\mathcal{L}(\mathcal {F})= \frac{1}{4b} \left(e^{-b \mathcal
{F}}-1\right),
\end{equation}
where  $b$ is the nonlinear parameter, $\mathcal
{F}=F_{\mu\nu}F^{\mu\nu}$, and $F^{\mu\nu}$ is the electromagnetic
field tensor. When $b\rightarrow 0$, the nonlinear electrodynamic
Lagrangian reduces to the linear Maxwell Lagrangian,
$\mathcal{L}(\mathcal {F})=-\mathcal {F}/{4}$. We further assume
the gauge and scalar fields have the following form \cite{Har1}
 \begin{eqnarray}
A_{\mu}=(\phi(r),0,0,0),  \   \    \    \psi=\psi(r),
\end{eqnarray}
By varying the Lagrangian (\ref{L}) with respect to the gauge
field $\phi$ and scalar field $\psi$, we obtain the field
equations,
\begin{eqnarray}\label{Phir}
\phi''(r)(1+4 b\phi'^2(r))+\frac{2}{r}\phi'(r)-\frac{2
\phi(r)\psi^2(r)}{f(r)} e^{-2b\phi'^2(r)} =0,
\end{eqnarray}
\begin{eqnarray}\label{psir}
\psi''(r)+\left(\frac{f'(r)}{f(r)}+\frac{2}{r}\right)\psi'(r)+\left(\frac{\phi^2(r)}{f^2(r)}-\frac{m^2}{f(r)}\right)\psi(r)=0.
\end{eqnarray}
It is clear that the nonlinear gauge field do not affect the
equation of $\psi$, however, the equation for $\phi$ will be
modified as given in (\ref{Phir}). When $b\rightarrow0$, Eq.
(\ref{Phir}) restores the corresponding equation of the standard
holographic superconductor in Maxwell theory \cite{Har1}. In order
to do an analytic calculations, we define a new variable
$z={r_{+}}/{r}$. In terms of this new variable, Eqs. (\ref{Phir})
and (\ref{psir}) become
\begin{eqnarray}\label{Phiz}
\phi''(z)\Bigg(1+4b \frac{z^4}{r_{+}^2}\phi'^2
\Bigg)+\frac{8bz^3}{r^2_{+}}\phi'^3(z)
-\frac{2\psi^2(z)}{z^2(1-z^3)}e^{-\frac{2bz^4}{r^2_{+}}\phi'^2(z)}
\phi(z)=0,
\end{eqnarray}
\begin{eqnarray}\label{psiz}
\psi''(z)-\frac{2+z^3}{z(1-z^3)}\psi'(z)+\Bigg[\frac{\phi^2(z)}{r_{+}^2(1-z^3)^2}-\frac{m^2}{z^2(1-z^3)}\Bigg]\psi(z)=0,
\end{eqnarray}
where we have also used
\begin{eqnarray}\label{fz}
f(z)=r_{+}^2 \left(\frac{1}{z^2}-z\right).
\end{eqnarray}
In order to avoid the divergence near the horizon, we assume the
boundary conditions $\phi(z=1)=0$ and
$\psi(z=1)=\frac{3}{2}\psi'(z=1)$. Also, on the boundary regime
where $z\rightarrow0$, the asymptotic solutions for Eqs.
(\ref{Phiz}) and (\ref{psiz}) are
 \begin{eqnarray} \label{BC.Phi}
\phi \approx \mu-\frac{\rho}{r_{+}}z,
 \end{eqnarray}
and
\begin{eqnarray}
\psi(z) \approx \frac{\psi_{-}}{r_{+}^{\Delta_{-}}}
z^{\Delta_{-}}+ \frac{\psi_{+}}{r_{+}^{\Delta_{+}}}z^{\Delta_{+}},
\end{eqnarray}
where
\begin{eqnarray}
\Delta_{\pm}=\frac{3\pm\sqrt{9+4 m^2}}{2},
\end{eqnarray}
and $\mu$ and $\rho$ are interpreted as the chemical potential and
charge density of the dual field theory \cite{Har1,Har2}. Also
$\psi_{\pm}=<\mathcal{O_{\pm}}>$ specifies the vacuum expectation
value of the dual operator $\mathcal{O_{\pm}}$ on the boundary.
According to the AdS/CFT correspondence, either $\psi_{+}$ or
$\psi_{-}$ will act as a condensation operator while the other
will act as a source \cite{Har1,Har2}. In the present work, we
choose $\psi_{-}=<\mathcal{O}_{-}>$ and $\psi_{+}$ as its source.
Since we would like the condensation to take place in the absence
of any source, we set $\psi_{+}=0$. Considering the fact that
$m^2$ must satisfy the Breitenlohner-Freedman (BF) bound
\cite{BF}, we set $m^2=-2$ \cite{Har2}.
\section{Analytical Investigation of the Holographic Superconductor\label{Field}}
In this section, we would like to analytically study the relation
between the critical temperature and charge density for the
holographic superconductor in the presence of EN electrodynamics.
We shall employ the SL variational method. At the critical
temperature, $T=T_{c}$, we have $\psi=0$ and hence Eq.
(\ref{Phiz}) reduces to
\begin{eqnarray}\label{PhiT0}
\phi''(z)\Bigg(1+4b \frac{z^4}{r_{+c}^2}\phi'^2 \Bigg)+\frac{8bz^3}{r^2_{+c}}\phi'^3(z)=0.
\end{eqnarray}
The above equation has the following solution,
 \begin{eqnarray}\label{Phibeta1}
\phi(z)=\int_{1}^{z}{dz \frac{r_{+c}}{2z^2
\sqrt{b}}\sqrt{L_{W}\left(\frac{4bz^4 \beta^2}{r_{+c}^2}\right)}},
\end{eqnarray}
where $\beta$ is a constant of integration and
$L_W(x)={LambertW(x)}$ is the Lambert function which satisfies
\cite{Lambert}
\begin{equation}
L_W(x)e^{L_W(x)}=x,
\end{equation}
and has the following series expansion
\begin{equation}\label{LW}
L_W(x)=x-x^2+\frac{3}{2}x^3-\frac{8}{3}x^4+....
\end{equation}
Clearly, series (\ref{LW}) converges for $|x| <1$. Expanding Eq.
(\ref{Phibeta1}) and keeping terms up to $\mathcal{ O} (b)$, we
find
 \begin{eqnarray}\label{Phibeta}
\phi(z)=-\beta(1-z)+\frac{2\beta^3 b}{5r_{+c}^2}
\left(1-z^5\right)+\mathcal{ O} (b^2),
\end{eqnarray}
Equating $\phi^{'}(z=0)$ from Eq. (\ref{Phibeta}) and Eq.
(\ref{BC.Phi}), we find $\beta=-\rho/r_{+c}$, where at $T=T_c$ we
have set $r_{+}=r_{+c}$. Substituting $\beta$ into Eq.
(\ref{Phibeta}), we arrive at
\begin{eqnarray}
\phi(z)&=&\frac{\rho}{r_{+c}}(1-z)-\frac{2\rho^3 b}{5 r_{+c}^5}
(1-z^5)
\nonumber\\&=&\frac{\rho}{r_{+c}}(1-z)\Bigg\{1-\frac{2\rho^2}{5r_{+c}^4}b
(1+z+z^2+z^3+z^4)\Bigg\} \nonumber\\
&=&\lambda
r_{+c}(1-z)\Bigg\{1-\frac{2}{5}b\lambda^2(1+z+z^2+z^3+z^4)\Bigg\},
\label{Phi}
\end{eqnarray}
with $b\lambda^2<1$,  and in the last step we have defined
\begin{eqnarray}\label{lam}
\lambda=\frac{\rho}{r^2_{+c}}.
\end{eqnarray}
It should be noted that we have investigated the effects of the
nonlinear corrections up to the leading order in the nonlinear
parameter $b$. In the next step, we start from the field equation
for the scalar field $\psi$ by writing it near the critical
temperatures $T\approx T_{c}$. Thus Eq. (\ref{psiz}) can be
rewritten as
\begin{eqnarray}\label{psiddprime}
&&\psi''(z)-\frac{2+z^3}{z(1-z^3)}\psi'(z)-\frac{m^2}{z^2(1-z^3)}\psi(z)\nonumber\\
&&+\frac{\lambda^2}{r_{+}^2(1+z+z^2)^2}
\left(1-\frac{2b\lambda^2}{5}\zeta(z)\right)\psi(z)=0,
\end{eqnarray}
where $\zeta(z)=1+z+z^2+z^3+z^4$, and we have used Eq. (\ref{Phi})
as well. In order to match the behavior at the boundary one
defines $\psi$ in the following form
\begin{eqnarray}\label{psi}
\psi(z)=\frac{<\mathcal{O}_{-}>}{\sqrt{2}r_{+}}zF(z),
\end{eqnarray}
where we introduce a trial function
\begin{eqnarray}\label{F}
F(z)=1-\alpha z^2,
\end{eqnarray}
which satisfies the boundary conditions $F(0)=1$ and
$F^{\prime}(0)=0$ near the boundary $z=0$ \cite{sup7}. Inserting
Eq. (\ref{psi}) into Eq. (\ref{psiddprime}), one can show that the
trial function satisfies the following second-order SL self
adjoint differential equation
\begin{eqnarray}
(1-z^3)F''(z)-3z^2F'(z)-zF(z)+\frac{\lambda^2(1-z)}{(1+z+z^2)^2}\left(1-\frac{4b\lambda^2}{5}\zeta(z)\right)F(z)=0.
\end{eqnarray}
The above equation is solved subject to the boundary condition
$F'(0)=0$. The expression for estimating the minimum eigenvalue of
$\lambda^{2}$ is provided by \cite{Gan}
\begin{eqnarray}\label{lambda2}
\lambda^2=\frac{\int_{0}^{1}dz\left[p(z)[F'(z)]^2+zF^2(z)\right]}
{\int_{0}^{1}dz q(z)[F(z)]^2},
\end{eqnarray}
where
\begin{eqnarray}
&&p(z)=1-z^3,   \\
&& q(z)=\frac{1-z}{1+z+z^2}\left(1-\frac{4}{5}
b\lambda^2\mid_{b=0}\zeta(z)\right).
\end{eqnarray}
Here we have performed a perturbative expansion $b \lambda^2$ and
retain only the terms that are linear in $b$ such that
\begin{eqnarray}
b \lambda^2=b \left(\lambda^2|_{b=0}\right) +\mathcal{O} (b^2),
\end{eqnarray}
where $\lambda^2|_{b=0}$ is the value of $\lambda^2$ for $b=0$. In
fact, we have only retain  the terms that are linear in nonlinear
parameter $b$. Having Eqs. (\ref{T}) and (\ref{lam}) in mind, the
critical temperature $T_{c}$ in terms of the charge density $\rho$
can be obtained as
\begin{eqnarray}\label{Tc}
T_{c}=\frac{3}{4\pi} \sqrt{\frac{\rho}{\lambda}}.
\end{eqnarray}
Our strategy for calculating the critical temperature for
condensation is to minimize the function (\ref{lambda2}) with
respect to the coefficient $\alpha$ and then substituting the
minimum value of $\lambda$ in Eq.(\ref{Tc}) we find the maximum
value of $T_c$. For $b=0$, one gets
\begin{eqnarray}
\lambda^2= \frac{10\alpha^2-6\alpha+6}{(2 \pi+4 \pi
\alpha)\sqrt{3}+(12\alpha+12 \alpha^2-6)\ln(3)-36\alpha-13
\alpha^2},
\end{eqnarray}
whose minimum is $\lambda^2_{\rm min}=1.2683$ at $\alpha=0.2389$.
For this value of $\lambda$, the critical temperature reads
$T_{c}=0.2250 \sqrt{\rho}$. For other values of the nonlinear
parameter $b$, we have summarized our results in table $1$.
\begin{center}
\begin{tabular}{|c|c|c|c|c|}
\hline
$b$ \quad &   $\alpha$\quad &   $\lambda^2_{\rm min}$\quad  &   $T_{c}$\quad \\
\hline
$0$ \quad &   0.2389 \quad &    1.2683\quad  &   0.2250 ${\sqrt{\rho}}$ \quad \\
\hline
$0.1$ \quad&   0.2449\quad &   1.4892\quad &   0.2161 ${\sqrt{\rho}}$ \quad\\
\hline
$0.2$ \quad&    0.2533\quad &   1.8028\quad &    0.2060 ${\sqrt{\rho}}$ \quad \\
\hline
$0.3$ \quad &   0.2660\quad &   2.2825\quad  &   0.1942 ${\sqrt{\rho}}$ \quad \\
\hline
\end{tabular}
\\[0pt]
Table $1$: The values of the eigenvalue $\lambda^2_{\rm min}$,
parameter $\alpha$  and the critical temperature $T_{c}$ for
several values of the nonlinear parameter $b$. \label{tab1}
\end{center}
We have compared in table $2$ the numerical and analytical results
for the critical temperature of holographic superconductors in the
presence of two kind of nonlinear gauge field, namely BI nonlinear
electrodynamics \cite{ZPCJ.NN, BIBR} and EN electrodynamics
\cite{ZPCJ.NN}. In this table, the analytical results for BI case
are taken from \cite{BIBR} (which also coincide with those
obtained in \cite{Gan}) and the numerical result for both BI and
EN are invoked from \cite{ZPCJ.NN}. We observe that for both
nonlinear electromagnetic models, the critical temperature $T_{c}$
decreases with increasing the nonlinear parameter $b$. Also for
the fixed values of the nonlinear parameter $b$, the critical
temperature $T_{c}$ for BI form is larger compared with the EN
form. This implies that in case of the EN electrodynamics, the
condensation is more difficult to create with respect to the BI
nonlinear electrodynamics. Our analytical results are also
consistent with the numerical results obtained in Ref.
\cite{ZPCJ.NN}. Indeed, from table $2$, we see that the values
obtained through both analytical and numerical results are in the
same order and the difference is very small. This difference may
be due to the perturbative technique, where we have only
considered terms up to the first order in the nonlinear coupling
parameter $b$. But in both of these approaches the critical
temperature, $T_{c}$, decreases with increasing the values of the
nonlinear parameter $b$.

\begin{center}
\begin{tabular}{|c|c|c|c|c|}
\hline $b$ \quad &   $T_{c}^{\rm EN}\mid_{Analytical}$\quad &
$T_{c}^{\rm EN}\mid_{Numerical}$\quad  &
$T_{c}^{\rm BI}\mid_{Analytical}$\quad  &   $T_{c}^{\rm BI}\mid_{Numerical}$\quad \\
\hline
$0$ \quad &   0.2250 $\sqrt{\rho}$ \quad &   0.2255$\sqrt{\rho}$\quad  &   0.2250$\sqrt{\rho}$\quad  &   0.2255$\sqrt{\rho}$\quad \\
\hline
$0.1$ \quad&   0.2161 $\sqrt{\rho}$\quad &  0.2247 $\sqrt{\rho}$\quad &  0.2228$\sqrt{\rho}$\quad &   0.2253$\sqrt{\rho}$\quad \\
\hline
$0.2$ \quad&   0.2060 $\sqrt{\rho}$\quad &   0.2225$\sqrt{\rho}$\quad &   0.2206$\sqrt{\rho}$\quad &   0.2247$\sqrt{\rho}$\quad \\
\hline
$0.3$ \quad&   0.1942 $\sqrt{\rho}$ \quad &   0.2217$\sqrt{\rho}$\quad &   0.2184$\sqrt{\rho}$\quad &   0.2237$\sqrt{\rho}$\quad \\
\hline
\end{tabular}
\\[0pt]
Table $2$: The critical temperature $T_{c}$ for different values
of the nonlinear parameter $b$ for EN and BI electrodynamics
obtained by the analytical SL  and numerical methods. \label{tab2}
\end{center}
\section{Critical Exponent and Condensation Values}\label{Cri}
In this section, we would like to calculate the condensation value
in the boundary field theory. We start from Eq. (\ref{Phiz}) by
writing it near the critical point. Substituting Eq. (\ref{psi})
into Eq. (\ref{Phiz}), we arrive at
\begin{eqnarray}\label{Phi.near T}
\phi''(z)\Bigg(1+4b \frac{z^4}{r_{+}^2}\phi'^2
\Bigg)+\frac{8bz^3}{r^2_{+}}\phi'^3(z) =\frac{\langle
\mathcal{O}_{-} \rangle ^2}{r_{+} ^2}\frac{F^2(z)}{1-z^3}
\Bigg[1-\frac{2bz^4}{r_{+}^2}\phi'^2(z)\Bigg]\phi(z)+\mathcal{O}
(b^2).
\end{eqnarray}
where we have used
\begin{eqnarray}
\exp\left(-\frac{2bz^4}{r_{+}^2}\phi'^2(z)\right)=1-\frac{2bz^4}{r_{+}^2}\phi'^2(z)+\mathcal{O}
(b^2).
\end{eqnarray}
Then, we expand $\phi(z)$ perturbatively in the small parameter
${\langle \mathcal{O} \rangle ^2}/{r_{+} ^2}$ as follows:
\begin{eqnarray}\label{Phi.bast}
\frac{\phi(z)}{r_{+}}=\lambda
(1-z)\Bigg\{1-\frac{2b\lambda^2}{5}\zeta(z)\Bigg\}+\frac{\langle
\mathcal{O}_{-} \rangle ^2}{r_{+} ^2}\chi(z)+... .
\end{eqnarray}
Substituting Eq. (\ref{Phi.bast}) into (\ref{Phi.near T}) in first
order with respect to the ${\langle \mathcal{O} \rangle ^2}/{r_{+}
^2}$, we obtain the following differential equation
\begin{eqnarray} \label{chi}
\chi''(z)[1+4b\lambda^2 z^4]+24b \lambda^2 z^3
\chi'(z)=\frac{\lambda
F^2(z)}{1+z+z^2}\Bigg\{1-\frac{2b\lambda^2}{5}(\zeta(z)+5z^4)\Bigg\},
\end{eqnarray}
where $\chi(z)$ satisfies the boundary condition
\begin{eqnarray}\label{xhi}
\chi(1)=\chi'(1)=0.
\end{eqnarray}
Multiplying both side of (\ref{chi}) by factor $(1+4b\lambda^2
z^4)^{1/2}$, we can write it as
\begin{eqnarray}\label{dxhi}
\frac{d}{dz}\Bigg[(1+4b\lambda^2 z^4)^{3/2}\chi'(z)\Bigg]=
(1+4b\lambda^2 z^4)^{1/2}\frac{\lambda
F^2(z)}{1+z+z^2}\Bigg\{1-\frac{2b\lambda^2}{5}(\zeta(z)+5z^4)\Bigg\}.
\end{eqnarray}
Using the boundary condition (\ref{xhi}) and integrating Eq.
(\ref{dxhi}) in the interval $[0,1]$, we finally obtain
\begin{eqnarray}
\chi'(0)=-\lambda\mathcal{A},
\end{eqnarray}
where
\begin{eqnarray}
\mathcal{A}=\int_{0}^{1}dz(1+4b\lambda^2 z^4)^{1/2}
\frac{F^2(z)}{1+z+z^2}\Bigg\{1-\frac{2b\lambda^2}{5}\left(\zeta(z)+5z^4\right)\Bigg\}.
\end{eqnarray}
Combining Eqs. (\ref{BC.Phi}) and (\ref{Phi.bast}), we may write
\begin{eqnarray}\label{murho}
\frac{\mu}{r_{+}}-\frac{\rho}{r_{+} ^2} z &=& \lambda (1-z)
\Bigg\{ 1-\frac{2b\lambda^2}{5} \xi(z) \Bigg\}+\frac{\langle
\mathcal{O}_{-} \rangle ^2}{r_{+} ^2} \chi (z)\nonumber\\&=&
\lambda (1-z) \Bigg\{ 1-\frac{2b\lambda^2}{5} \xi(z)
\Bigg\}+\frac{\langle \mathcal{O}_{-} \rangle ^2}{r_{+} ^2}
\left(\chi(0)+z \chi'(0)+...\right),
\end{eqnarray}
where in the last step we have expanded $\chi(z)$ around $z = 0$.
Now, if we equate the coefficients of $z$ on both sides of Eq.
(\ref{murho}), we get
\begin{eqnarray}\label{RL}
\frac{\rho}{r_{+} ^2}=\lambda-\frac{\langle \mathcal{O}_{-}
\rangle ^2}{r_{+} ^2}\chi'(0)= \lambda \left(1+\frac{\langle
\mathcal{O}_{-} \rangle ^2}{r_{+} ^2} \mathcal{A} \right).
\end{eqnarray}
Substituting $\lambda={\rho}/{r^2_{+c}}$  into Eq. (\ref{RL}), and
using the definition for the temperature given in (\ref{T}), we
finally get the expression for the order parameter $\langle
\mathcal{O}_{-} \rangle$ near the critical temperature $T_{c}$ as
\begin{eqnarray}\label{CE}
\langle \mathcal{O}_{-} \rangle=\gamma T_{c}
\sqrt{1-\frac{T}{T_{c}}},
\end{eqnarray}
where
\begin{eqnarray}
\gamma=\frac{4\pi\sqrt{2}}{\sqrt{3\mathcal{A}}}.
\end{eqnarray}
From Eq. (\ref{CE}) we find that the critical exponent of the
system always takes the mean field value $1/2$. This result is
independent of the nonlinear parameter $b$. Also, it is clear that
$\langle \mathcal{O}_{-} \rangle$ is zero at $T = T_{c}$ and
condensation occurs for $T < T_{c}$. In table $3$, we present the
condensation values $\gamma$ obtained for a $(2+1)$-dimensional
holographic superconductor, by the analytical SL method for both
BI and EN electrodynamics and for different values of the
nonlinear parameter $b$  Again, the analytical results for BI case
are taken from \cite{Gan}. In both cases, we observe that the
condensation value $\gamma$ increases with increasing the
nonlinear parameter $b$. This indicates that the higher nonlinear
electrodynamics corrections to the gauge field make the
condensation to be formed harder. On the other hand, comparing
these two types nonlinear electrodynamics, we find that the
condensation operator for the EN is larger than that of BI which
means that the EN electrodynamics makes the condensation harder
with respect to the BI nonlinear electrodynamics.
\begin{center}
\begin{tabular}{|c|c|c|c|c|}
\hline
$b$ \quad &   $\gamma ^{\rm EN}$\quad &   $\gamma^{\rm BI}$\quad \\
\hline
$0$ \quad &   8.07 \quad &    8.07\quad \\
\hline
$0.1$ \quad&   8.53\quad &  8.19\quad \\
\hline
$0.2$ \quad&   9.16\quad &   8.33\quad \\
\hline
$0.3$ \quad &   10.03\quad &    8.54\quad \\
\hline
\end{tabular}
\\[0pt]
Table $3$: A comparison of the analytical results for the
expectation value of the condensation operator, $\gamma$, for EN
and BI electrodynamics. \label{tab3}
\end{center}
\section{Concluding remarks\label{Con}}
In the present paper, we analytically studied the properties of
the holographic $s$-wave superconductors in the presence of
exponential correction to the Maxwell electrodynamics in the
background of Schwarzschild AdS black holes. Adopting the probe
limit, where the scalar and gauge field do not back react on the
metric background, we found the nonlinear coupling parameter
influences the critical temperature and the condensation operator
near the critical point. We performed our calculations up to the
first order in the nonlinear coupling parameter $b$. We observed
that the critical temperature $T_{c}$ decreases with increasing
the nonlinear parameter $b$, which means that the condensation
become harder to be formed for higher nonlinear corrections. This
result is similar to the holographic $s$-wave superconductors with
BI nonlinear electrodynamics \cite{Gan}. However, we found that
the EN electrodynamics has stronger effect on the condensation
formation compared to the BI electrodynamics.  The analytical
results obtained in this work are in good agreement with the
existing numerical results for the holographic superconductors
with EN electrodynamics \cite{ZPCJ.NN}. The critical exponent of
the condensation also comes out to be $1/2$ which is the universal
value in the mean field theory. Thus, the superconductor phase
transition appeared in the linear Maxwell electrodynamics also
exists in its nonlinear extension. Our work helps to give a better
understanding of the holographic superconductors  when the gauge
field is in the form of EN electrodynamics.

Finally, we would like to mention that in this work we only
considered the probe limit. It would be interesting if one could
generalize this study away from the probe limit, i.e., by taking
the backreaction of scalar and gauge fields on the metric
background into account, and analytically discuss the effects of
both backreaction and the nonlinear coupling parameters on the
properties of the holographic superconductors. It is also worthy
to extend this investigation to higher curvature corrections
theories such as Gauss-Bonnet gravity. These issues are now under
investigation and the results will be appeared in the near future.
\acknowledgments{We thank Shiraz University Research Council. This
work has been supported financially by Research Institute for
Astronomy and Astrophysics of Maragha (RIAAM), Iran.}

\end{document}